\documentclass[%
 aip,
 amsmath,amssymb,
reprint, onecolumn %%%%% Comment onecolumn to override one column format %%%%%%%
]{revtex4-1}

\usepackage{graphicx}% Include figure files
\usepackage{dcolumn}% Align table columns on decimal point
\usepackage{bm}% bold math
\usepackage[utf8]{inputenc}
\usepackage[T1]{fontenc}
\usepackage{newtxmath}
\usepackage{etoolbox}

\makeatletter
\def\@email#1#2{%
 \endgroup
 \patchcmd{\titleblock@produce}
  {\frontmatter@RRAPformat}
  {\frontmatter@RRAPformat{\produce@RRAP{*#1\href{mailto:#2}{#2}}}\frontmatter@RRAPformat}
  {}{}
}%
\makeatother
\begin{document}

\preprint{AIP/123-QED}

\title[Direct density estimation using muon radiography]{Direct density estimation using muon radiography}
% Force line breaks with \\
\author{J. Peña-Rodríguez}
 \email{penarodriguez@uni-wuppertal.de}
 \affiliation{Fakultät für Mathematik und Naturwissenschaften, Bergische Wuppertal Universität, Wuppertal, Germany}

\date{\today}% It is always \today, today,

\begin{abstract}

Muography or muon radiography estimates the density distribution of natural or anthropic structures by measuring the traversing flux of atmospheric muons. Muography has been implemented to image volcanoes, glaciers, tunnels, line-shores, pyramids, and dams. The mass variation of the structure is deduced by the ratio between the open-sky muon flux and the target traversing flux, resulting in a relative measurement of the density. We present a novel method for directly measuring the average density of the target. The methodology uses known muography variables such as open-sky and traversing muon flux, incident zenith angle, and muon path length. We validated the method with muography-simulated data from iron, aluminum, standard rock and water phantoms, as well as real data from the Khufu pyramid, the Eiger Glacier, and the Canfranc Underground Laboratory. 

\end{abstract}

\maketitle

\section{\label{muography} Introduction}
Muon radiography uses atmospheric muon flux to scan the inner density distribution of natural or human-made structures. The muon flux starts at the top of the Earth atmosphere as a result of the decay of kaons ($K^{\pm} \rightarrow \mu^{\pm} + \nu_{\mu}(\bar{\nu}_{\mu})$ [$\sim63.55\%$] $\&$ $K^{\pm} \rightarrow \pi^{\pm} + \pi^0$ [$\sim20.66\%$]) and pions ($\pi^{\pm} \rightarrow \mu^{\pm} + \nu_{\mu}(\bar{\nu}_{\mu})$ [$\sim100\%$]). The muon flux at ground level follows a $\cos^2 \theta$ distribution, where $\theta$ is the zenith angle, and the vertical muon flux is up to two orders of magnitude stronger than the horizontal flux. 

Muons interact weakly with matter, losing energy mainly by ionization or bremmstrahlung. This property allows muons to traverse large natural or anthropic structures and use them as an scanning tool. Muography has been applied to monitor pyramids \cite{Procureur2023, Morishima2017}, volcanoes \cite{Tioukov2019,Olh2019,DErrico2020,RosasCarbajal2017,Tanaka2018,VesgaRamrez2021,PeaRodrguez2020}, nuclear reactors \cite{Procureur2023_nuc}, glaciers \cite{Nishiyama2017,Ariga2018,Nishiyama2019}, tunnels \cite{Mao2023,Thompson2020}, mines \cite{Liu2024,Borselli2022}, and dams \cite{Olh2023,LzaroRoche2021}.

The muon flux attenuation after traversing the target can be modeled by the Lambert-Beer law,

\begin{equation}
    I = I_0 e^{-\int_L \mu dx} ,
\end{equation}

where $I$ is the traversing muon flux, $I_0$ is the open sky muon flux, $\mu$ is the linear attenuation coefficient, and $L$ is the muon path length. The relative flux attenuation is,

\begin{equation}
    \ln \frac{I}{I_0} = - \int_L \mu dx ,
\end{equation}

where the relation $I/I_0$ is defined as transmittance. The linear attenuation coefficient is directly related with the material density as

\begin{equation}
    \mu = \kappa \rho, 
\end{equation}

where $\kappa$ is the mass attenuation coefficient in [cm$^2$g$^{-1}$] and $\rho$ is the density of the material. Then, we define the opacity from the Lambert-Beer law as

\begin{equation}
   \varrho =  -\frac{1}{\kappa}\ln T = \int_L  \rho dx.
\end{equation}

Assuming an average density of the material along the muon path, the opacity is

\begin{equation}
    \varrho = - \frac{1}{\kappa} \ln T = \bar{\rho} L.
\end{equation}

Today, muography estimates relative changes in target density, and in some cases a known density sample of the scanned target is used as a reference to obtain the average density in the rest of the muon flux trajectories \cite{Lesparre2012}. In this work, we propose a novel method for estimating the average density along the muon path directly from muography.

\section{\label{concept} Density estimation principle}

The measured muon flux passing through a given structure is the integration of the differential muon flux $\Phi(E, \theta)$ from the minimum crossing energy $E_{min}$ to the infinite. The open-sky muon flux for a given incident zenith angle is,

\begin{equation}
    I_0(\theta) = \int_0^{\infty} \Phi(E, \theta) dE  \ \ [cm^{-2}sr^{-1}s^{-1}],
\end{equation}

and the muon flux crossing a target,

\begin{equation}
    I(\theta, \varphi) = \int_{E{min}}^{\infty} \Phi(E, \theta) dE.
\end{equation}

\begin{figure}[h!]
    \centering
    \includegraphics[width=10cm]{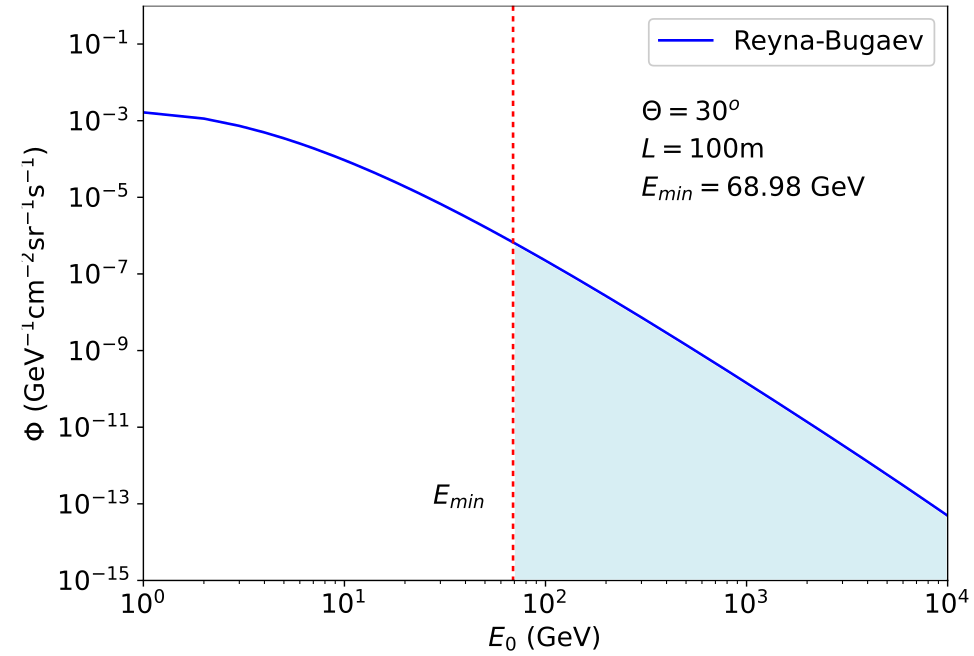}
    \includegraphics[width=5.5cm]{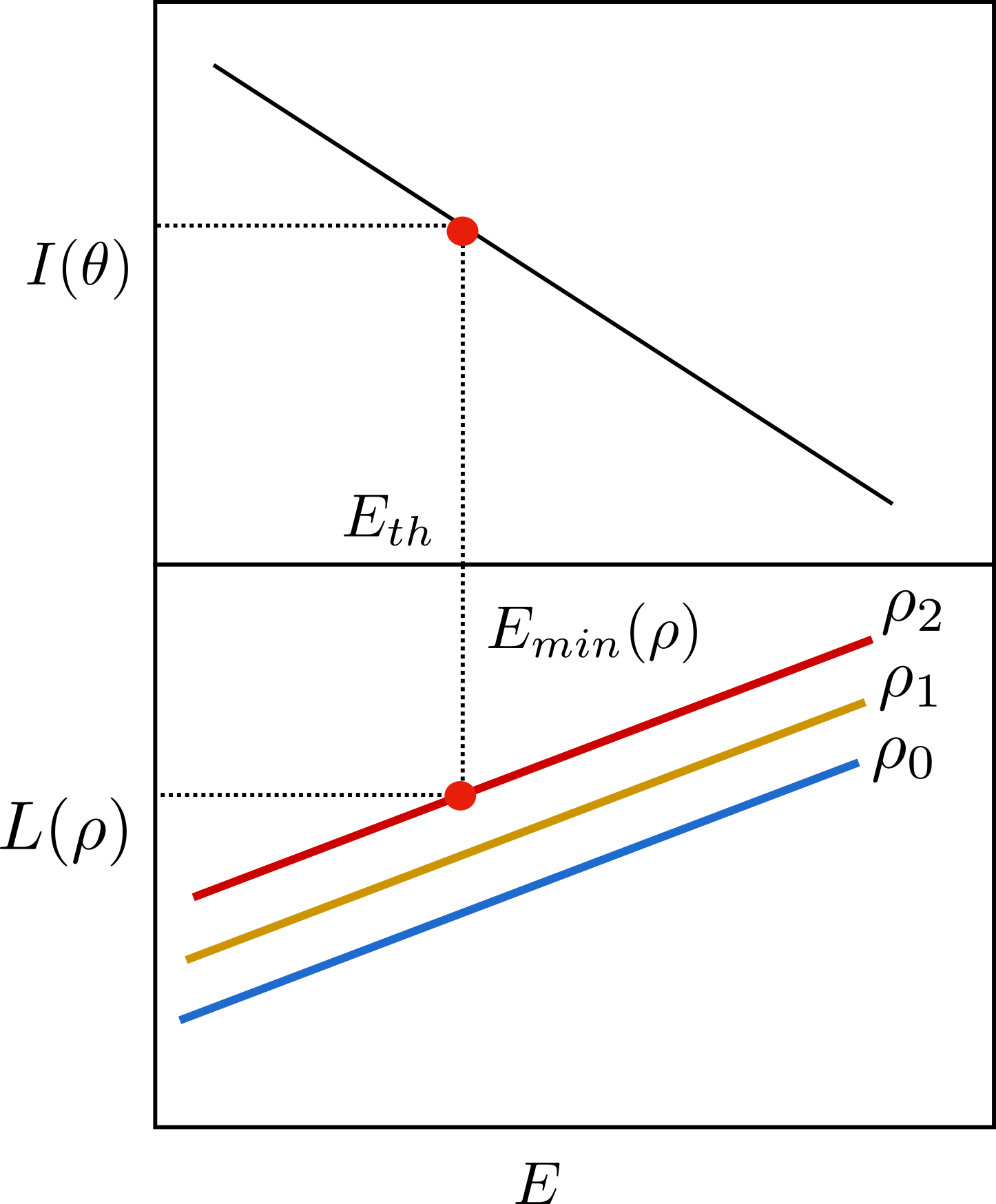}
    \caption{Differential muon flux integration. The shadow area represents the traversing muon flux along 100\,m of standard rock at 30\,deg zenith (left). Direct density estimation principle (right).}
    \label{fig:flux_integration}
\end{figure}

The observed muon flux after crossing the target also depends on the azimuth angle $\varphi$ because the target material budget varies for all ($\theta, \varphi$) combinations. The missed muon flux can be obtained by subtraction of the observed flux from the open-sky flux.

\begin{equation}
    I_{miss}(\theta, \varphi) = \int_0^{E{min}} \Phi(E, \theta) dE = I_0(\theta) - I(\theta, \varphi).
\end{equation}

We can deduce the muon threshold energy by knowing the muon spectrum from a given zenith angle and integrating it from a threshold energy $E_{th}$ to the infinite, as shown in Figure \ref{fig:flux_integration}-left. By using the obtained curve ($I(E_{th}, \theta)$ vs. $E$) and the measured flux $I_{obs}$, we can estimate the muon minimum energy by,

\begin{equation}
    E_{min} = \operatorname*{argmin}_{E_{th}} \left[ I_0(E_{th}, \theta) - I(\theta, \varphi) \right]^2.
\end{equation}

Having $E_{min}$ and knowing the muon path length $L$, the material density can be estimated from the $I_{obs}(\theta)$ vs. path length vs. density space as shown in Figure \ref{fig:flux_integration}-right.

We also have to take into account the muon detector working principle. The detector records muon counting $N$ for a given zenith angle. The detected muon flux is estimated as,

\begin{equation}
    I(\theta, \varphi) = \frac{\eta(\theta, \varphi) N(\theta, \varphi)}{\mathcal{T}(\theta, \varphi)T},
\end{equation}
where $\mathcal{T}$ is the detector acceptance for a given ($\theta,  \varphi$) combination, $T$ the observation time, and $\eta$ the detector efficiency. 

% Then,
% \begin{equation}
%     E_{min} = \operatorname*{argmin}_{E_{min}} \left[ I(E_{th}, \theta) - \frac{\eta (\theta) N(\theta)}{\mathcal{T}(\theta)T} \right]^2 .
% \end{equation}

The errors of the proposed methodology come mainly from the muon flux estimation, the muon path length estimation, and the detector angular resolution. Finally, the average density of the material is estimated from a modified Nagamine model\cite{Nagamine1995},

\begin{equation}
    \rho = \frac{300\times 10^3}{L} \ln (1.5E+1),
\end{equation}
where $\rho$ in gcm$^{-3}$, $L$ in cm, and $E$ in TeV.

\section{\label{flux} Muon flux model}

We simulated the muon differential flux by means of the MUYSC code\cite{PeaRodrguez2024} running the Reyna-Bugaev model\citep{Reyna2006}. In the Reyna-Bugaev model the muon flux is,

\begin{equation}
    \Phi_R(p, \theta) = A_R \hat{p}^{-(a_3y^3+a_2y^2+a_1y+a_0)} \cos^3 \theta,
\end{equation}

where $p$ is the muon momentum in (GeV/c), $\theta$ is the incidence zenith angle,
$y = \log_{10} \hat{p}$ and $\hat{p} = p \cos \theta$. The fitting parameters are $A_R = 0.00253$, $a_0 = 0.2455$, $a_1 = 1.288$, $a_2 = -0.2555$, and $a_3 = 0.0209$.

\section{Method validation}

\subsection{Simulations}

We performed muon flux simulations for toluene (0.86\,gcm$^{-3}$), paraffin (0.93\,gcm$^{-3}$), water (1.0\,gcm$^{-3}$), standard rock (2.65\,gcm$^{-3}$), aluminum oxide (3.97\,gcm$^{-3}$), and iron (7.87\,gcm$^{-3}$) phantoms. The phantom is a cilinder 10\,m height and 6\,m diameter. The observation point is located at 30\,m from the cylinder center, with an elevation angle of 33\,deg, an aperture of -14.3\,deg $\leq \theta_{x,y} \leq 14.3$\,deg. The muon flux was simulated at sea level, where the telescope $\theta_{x,y}=0$ is equivalent to 57\,deg zenith.

\begin{figure}[h!]
    \centering
    \includegraphics[width=3cm]{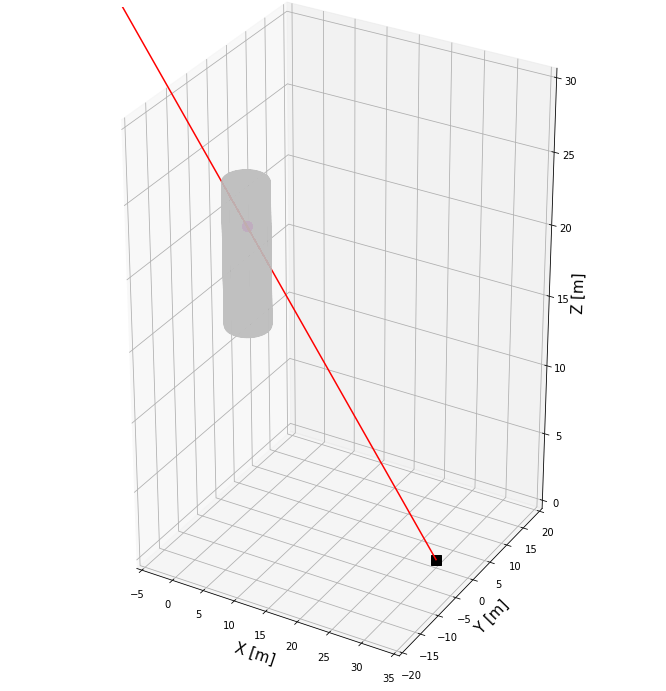}\hspace{1.2cm}
    \includegraphics[width=4cm]{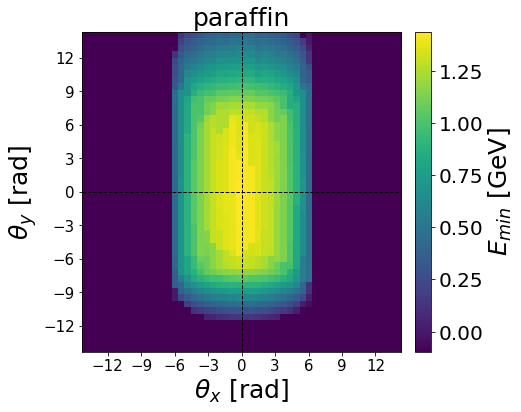}
    \includegraphics[width=4cm]{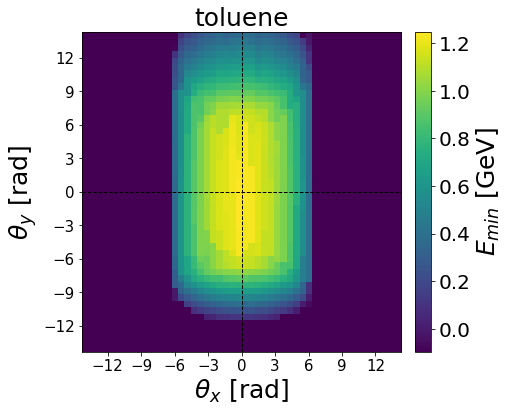}
    \includegraphics[width=4cm]{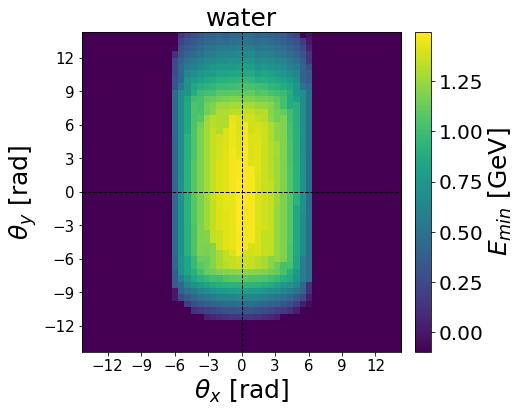}\\
    \includegraphics[width=4cm]{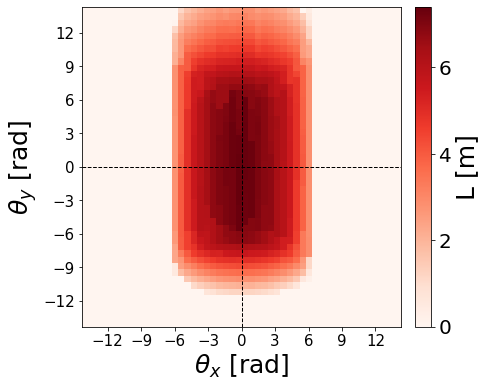}
    \includegraphics[width=4cm]{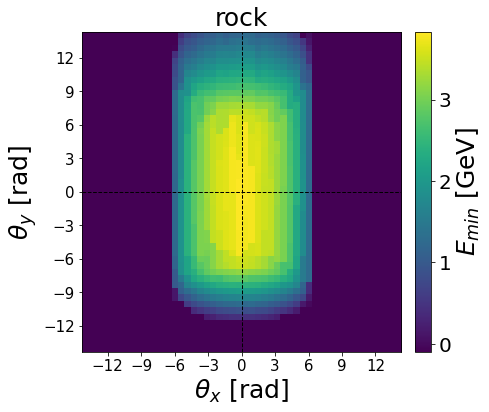}
    \includegraphics[width=4cm]{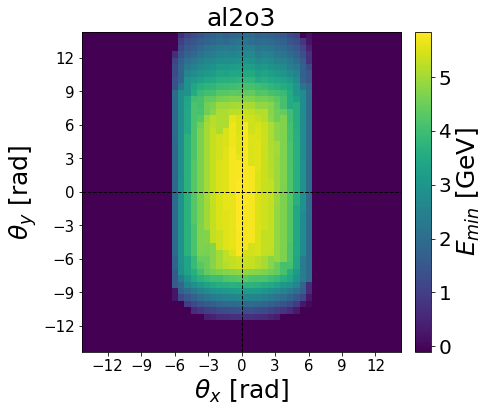}
    \includegraphics[width=4cm]{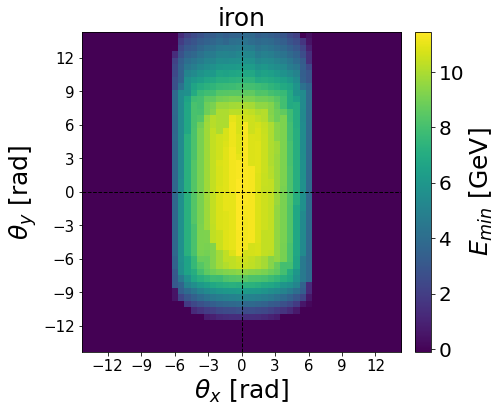}
    \caption{Simulated cylinder phantom (top-left). Path length histogram of the cylinder with dimensions 10\,m height and 6\,m diameter (bottom-left). Muon minimum energy required for traversing the phantom depending on the path length and the material (paraffin, toluene, water, standard rock, aluminum oxide, and iron).}
    \label{fig::phantom}
\end{figure}

\begin{figure}[h!]
    \centering
    \includegraphics[width=7.5cm]{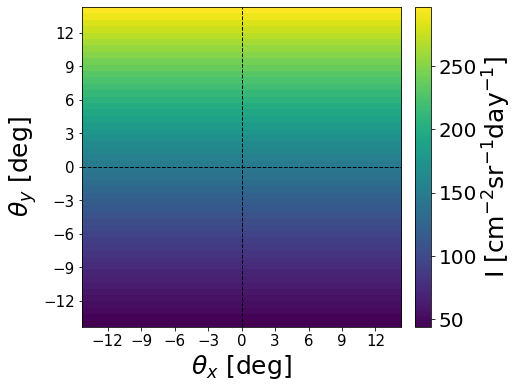}
    \includegraphics[width=7.5cm]{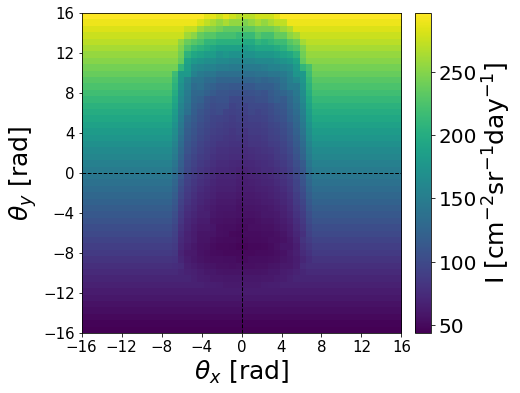}
    \caption{Open-sky (left) and traversing muon flux across the standard rock phantom (right).}
    \label{fig::phantom_std}
\end{figure}

\begin{figure}[h!]
    \centering
    \includegraphics[width=7.5cm]{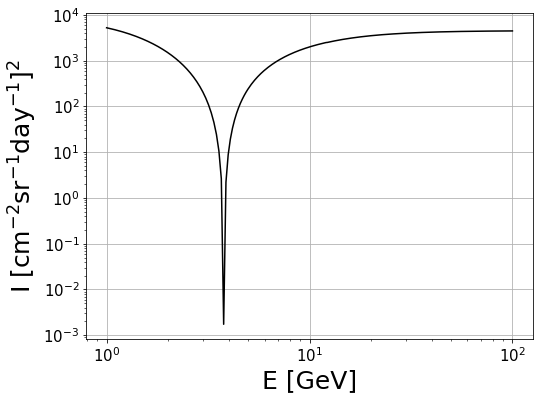}
    \includegraphics[width=7.5cm]{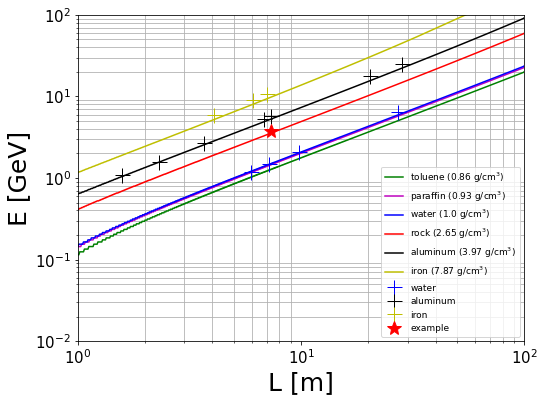}
    \caption{Energy required for traversing 7.36\,m of standard rock at 57\,deg zenith (left). Muon energy vs. path length for different phantoms of toluene (green), paraffin (cyan), water(blue), standard rock (red), aluminum (black), and iron (yellow) (right). Crosses represent the model prediction. The red star shows the example described in the text (57\,deg zenith, 67.34\,cm$^{-2}$sr$^{-1}$day$^{-1}$ traversing muon flux, 143.4\,cm$^{-2}$sr$^{-1}$day$^{-1}$ open-sky muon flux, and 7.36\,m path length).}
    \label{fig::phantom_dens}
\end{figure}

The muon energy loss in the phantom material was simulated with a parametric model implemented in the MUYSC code\cite{Lesparre2010, PeaRodrguez2024} and based on the Groom data tables\cite{GROOM2001}. Figure \ref{fig::phantom_std} shows the traversing muon flux for the standard rock phantom. At $\theta_{x,y}=0$ (57\,deg zenith) the traversing muon flux is 67.34\,cm$^{-2}$sr$^{-1}$day$^{-1}$ with a path length of 7.36\,m, while the open-sky muon flux is 143.4\,cm$^{-2}$sr$^{-1}$day$^{-1}$. The proposed method found a threshold muon energy of 3.77\,GeV as is shown in Figure \ref{fig::phantom_dens} left.

The simulated data fit accurately with the expected E vs. L curves for the different material as shown in Figure \ref{fig::phantom_dens}-right, taking into account that we know exactly the input parameters (the open-sky and traversing flux, the path length, and the incident zenith angle), the method operates with high accuracy. The estimated density values of the simulated phantoms are summarized in Table \ref{tab:sim_data}. 

\begin{table}[h!]
\caption{\label{tab:sim_data}Density estimation of the simulated phantom}
\begin{ruledtabular}
\begin{tabular}{lccc}
Material & Estimated density [gcm$^{-3}$] & Expected density [gcm$^{-3}$]  & Estimated/Expected ratio  \\ \hline
Toluene    & 0.86 & 0.86  & 1.00 \\ \hline
Paraffin & 0.97 & 0.93 & 1.05 \\ \hline
Water    & 1.01 & 1.00  & 1.01  \\ \hline
Standard rock    & 2.45 & 2.65  & 0.92  \\ \hline
Aluminum oxide    & 3.67 & 3.97  & 0.92  \\ \hline
Iron    & 7.79 & 7.87  & 0.99  \\ 
\end{tabular}
\end{ruledtabular}
\end{table}

The estimated density has a maximum error of about 10$\%$ for iron, $\sim 7.5\%$ for standard rock and aluminum oxide, and $< 5\%$ for water, paraffin and toluene.

\subsection{Data}

The density estimation method was tested with real muography data from the Khufu pyramid (Egipt) \cite{Procureur2023}, the Eiger glacier (Switzerland, European Alps)\cite{Nishiyama2019}, and the Canfranc Underground Laboratory\cite{Trzaska2019}. In those cases of study, we lack open-sky data, but we used the Reyna-Bugaev model for estimating it. The detected open-sky muon flux can vary slightly from the modeled flux depending on several factors, i.e. the detector calibration. We performed a correction of the modeled open-sky flux as follows:

\begin{equation}
    E_{min} = \operatorname*{argmin}_{E_{th}} \left[ \alpha I_0(E_{th}, \theta) - I(\theta, \varphi) \right]^2.
\end{equation}

where $\alpha$ is a scaling factor defined as the ratio measured/modeled open-sky flux at a given zenith angle in the observation site.

\subsubsection{Khufu pyramid}

The Great Pyramid is the largest of the three main pyramids in Egypt with a height of 146 meters and served as the tomb of Khufu. The Nagoya University muography group installed six muon detectors (EM1, EM2N, EM2C, EM2S, EM3, and EM4) made of nuclear emulsion films in the Khufu pyramid to look for hidden chambers. The detectors had an angular range of $|\tan \theta_{x,y}| \leq 1.0$ (45\,deg). We used the flux recorded by the EM1 detector because of its small uncertainty. The EM1 took data during 172\,days recording 9.48$\times 10^7$ tracks.

\begin{figure}[h!]
    \centering 
    \includegraphics[width=7cm]{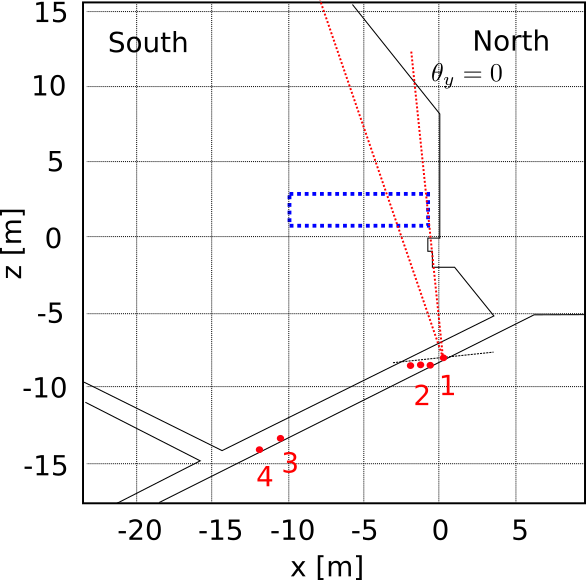}\hspace{0.2cm}
    \includegraphics[width=7cm]{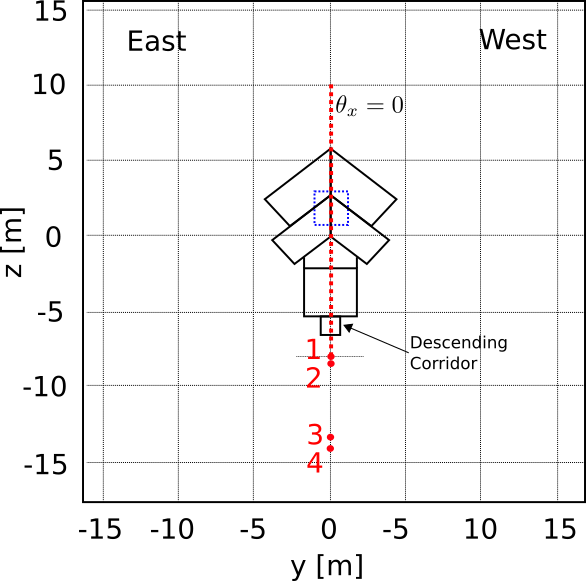}
    \caption{Location of the detectors EM1, EM2N, EM2C, EM2S, EM3, and EM4 inside the Khufu pyramid.}
    \label{fig::khufu}
\end{figure}

\begin{figure}[h!]
    \centering 
    \includegraphics[width=7cm]{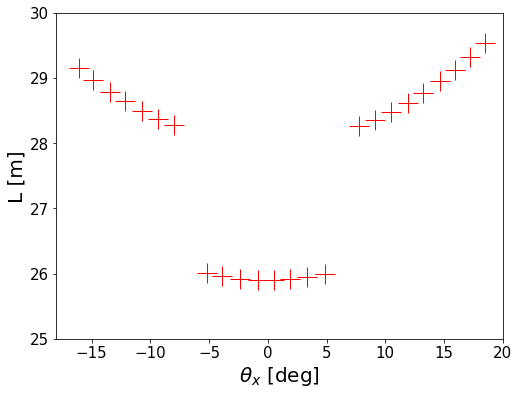}\hspace{0.2cm}
    \includegraphics[width=7cm]{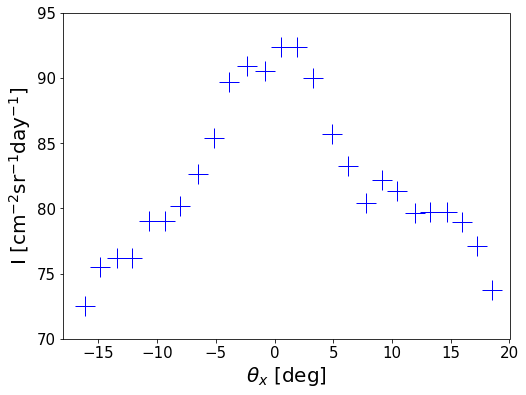}
    \caption{Estimated muon path length in rock (left) and measured muon flux at the EM1 observation point (right).}
    \label{fig::khufu_flux}
\end{figure}

Despite the fact that the main goal of the campaign was to explore hidden voids in the pyramid structure, our aim is to estimate the density of the rock around the detector EM1. Figure \ref{fig::khufu} shows the configuration of the emulsion detectors inside the Khufu pyramid and the location of the hidden chamber. Figure \ref{fig::khufu_flux} shows the muon path (left) and the measured muon flux (right).

The EM1 detector is located almost perpendicularly to the zenith 0 covering the same zenithal aperture for $\theta_x$ and $\theta_y$. In general, the muon flux varies with the zenith angle, but remains constant with the azimuthal angle. EM1 measures the muon flux in the range $-26.5$\,deg $ \leq  \theta_x < 26.5$\,deg in the region of $-14$\,deg $ \leq  \theta_y < -12$\,deg.

The dimensions of the chamber obtained by the monitoring campaign establish $2.18 \pm 0.17$\,m height , $2.02 \pm 0.06$\,m width, and $9.06 \pm 0.07$\,m length. The EM1 point is located $\sim 7.7$\,m below the base of the chamber and the slope of the external face of the pyramid is $\sim 51$\,deg. We calculated the path length of the muons that crossed the pyramid rock using the structural data described above as shown in Figure \ref{fig::khufu_flux} left. The measured muon flux shows an enhancement in the area ($|\theta_x| \lesssim 5$\,deg where the discovered chamber is located, as shown in Figure \ref{fig::khufu_flux} right.

The calibration parameter $\alpha=1.94$ was calculated taking into account that the open-sky muon flux at the zenith 0\,deg is $\sim$1.2$\times^3$\,cm$^{-2}$sr$^{-1}$s$^{-1}$ according to the observation point G1 (Alhazen) of a previous expedition \cite{Morishima2017}. 

\begin{figure}[h!]
    \centering
    \includegraphics[width=8.5cm]{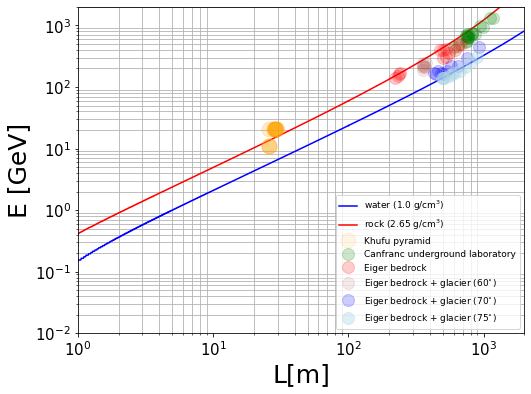}
    \includegraphics[width=8.5cm]{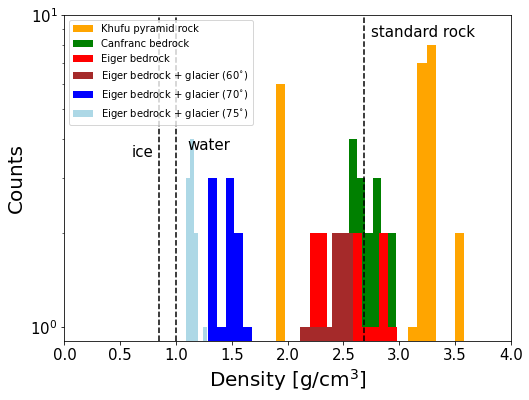}
    \caption{Method validation with data from the Khufu pyramid rock (yellow dots), the Eiger glacier (blue/brown dots), and the Canfranc underground laboratory (green dots). The expected density of standard rock and water are represented by the red and blue solid line respectively (left). Density distribution of the the Khufu pyramid rock, the Eiger glacier, and the bedrock of the Canfranc underground laboratory (right).}
    \label{fig::density}
\end{figure}

The minimum muon energy necessary, to cross the different muon paths along the pyramid rock, is overlayed on the standard rock curve in the curve $E$ vs. $L$, as shown in Figure \ref{fig::density} left. The average rock density of the pyramid was estimated to be 2.93\,gcm$^{-3}$, $10\%$ higher than the reported, as shown in Table \ref{tab:table2}.

\subsubsection{Canfranc Underground Laboratory}

\begin{figure}[h!]
    \centering
    \includegraphics[width=8.5cm]{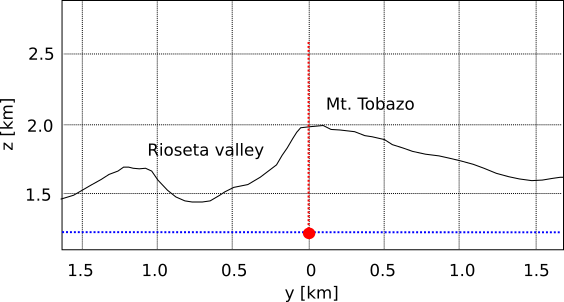}
    \hspace{2cm}
    \includegraphics[width=3.5cm]{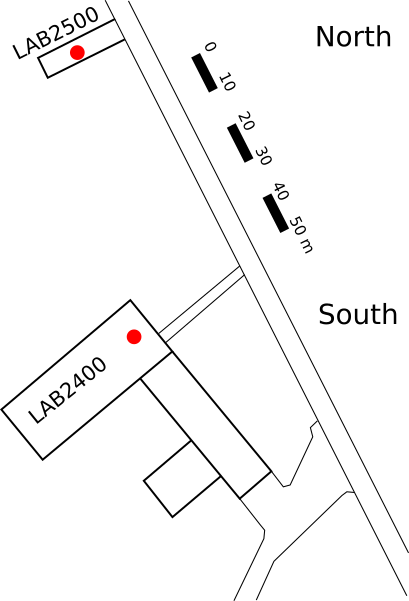}
    \caption{Location of the Canfranc underground laboratory at the Mount Tobazo (left). Location of the muon detector at the LAB2400 (1204.48\,m a.s.l.) and LAB2500 (1206.47\,m a.s.l.)(right).}
    \label{fig::canfranc}
\end{figure}

The Canfranc Underground Laboratory is located under the Mount Tobazo (Aragonese Pyrenees, Spain) providing a perfect shielding for research in neutrino and dark matter physics. In Canfranc there is a dedicated muon monitor composed of three layers of scintillation modules with a detection area of 1\,m$^2$ and covering all azimuth angles and zenith angles up to 80\,deg \cite{Trzaska2019}.

\begin{figure}[h!]
    \centering    
    \includegraphics[width=8cm]{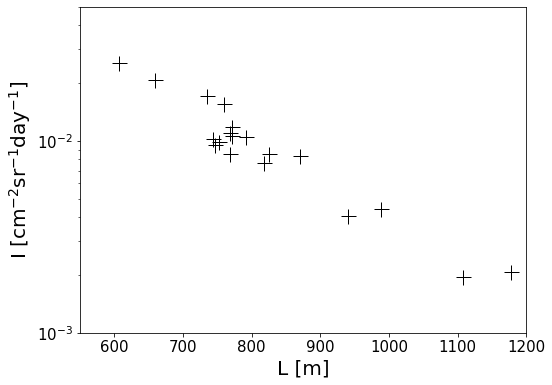}
    \caption{Measured muon flux depending on the muon path length at 40\,deg zenith at LAB2400 and LAB2500.}
    \label{fig::canfranc_flux}
\end{figure}

The muon detector was installed in two different places, the LAB2400 and LAB2500. The path length traversed by muons arriving from a given direction was determined using topographic data from the Advanced Land Observing Satellite (ALOS) (April 2018 release). Figure \ref{fig::canfranc} shows the location and distribution of the underground laboratory of Canfranc.

The maximum muon flux (25.5 $\times10^{-3}$\,cm$^{-2}$sr$^{-1}$day$^{-1}$) was observed in the direction of the Rioseta valley (40\,deg zenith and 150\,deg azimuth). Figure \ref{fig::canfranc_flux} shows the muon flux depending on the path length recorded from the LAB2400 and LAB2500 inside the underground facility. The estimated density of the bedrock by the authors was 2.73\,gcm$^{-3}$. We estimated a bedrock density of 2.69\,gcm$^{-3}$ as shown in Table \ref{tab:table2}.

\subsubsection{Eiger glacier}

\begin{figure}[h!]
    \centering
    \includegraphics[width=7cm]{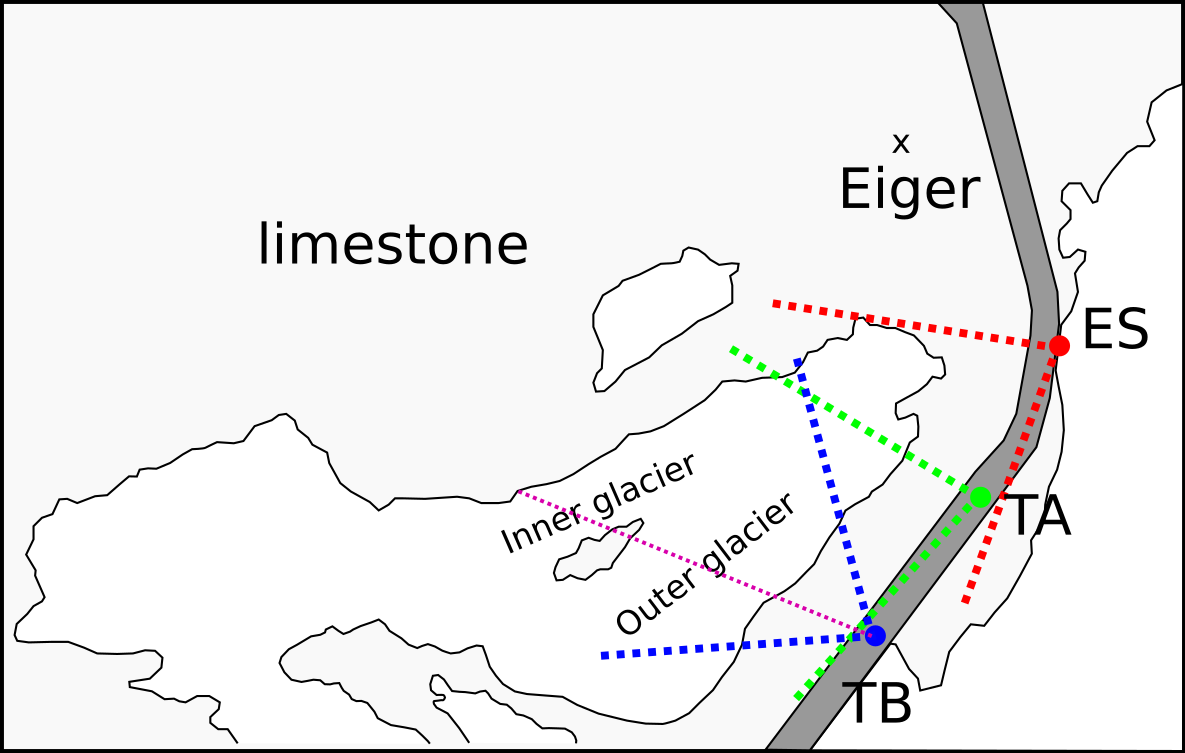}\hspace{0.5cm}
    \includegraphics[width=8cm]{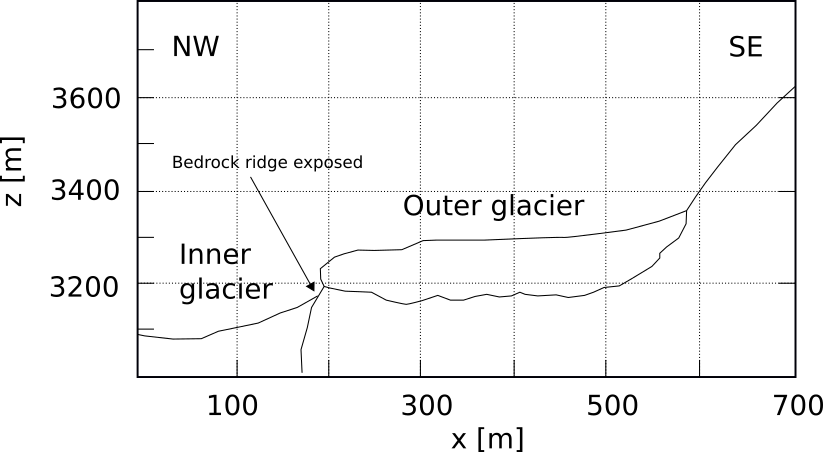}
    \caption{Top view of the Eiger glacier. Location of the monitoring stations ES (Eismeer station) 3159.9\,m, TA (Tunnel site A) 3186.4\,m, and TB (Tunnel site B) 3215.8\,m (left). Cross section of the Eiger glacier from the tunnel site B.}
    \label{fig::eiger}
\end{figure}

In 2019 muon radiography was used on the flank of Mt. Eiger (Switzerland, European Alps) to reconstruct the shape of the bedrock and the Eiger glacier. Particle detectors made of nuclear emulsion films were installed in three different places: ES (Eismeer station) 3159.9\,m, TA (Tunnel site A) 3186.4\,m, and TB (Tunnel site B) 3215.8\,m as shown in Figure \ref{fig::eiger}-left \cite{Nishiyama2019}. The bulk density of the bedrock was estimated from 16 rock samples and the mean value of  2.68 g/cm$^3$. They determined the dimensions of the glacier in terms of the ice-bedrock proportion along the muon trajectories from each detector, and then the results were combined to get a three-dimensional representation. Figure \ref{fig::eiger}-right shows the cross section of the Eiger glacier from the point of view of the TB detector.

\begin{figure}[h!]
    \centering
    \includegraphics[width=9cm]{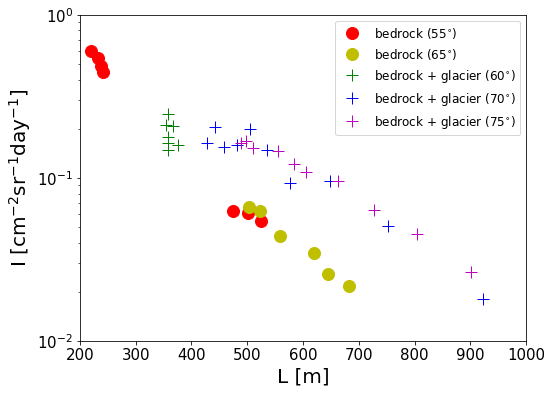}
    \caption{Muon flux crossing the Eiger glacier. The solid circles indicate the muon flux that traverses only the bedrock (TA station) while the crosses show the muon flux crossing the glacier and the bedrock (TB station).}
    \label{fig::eiger_flux}
\end{figure}

The muon flux that traverses the glacier increases drastically in comparison to that of crossing pure bedrock, as shown in Figure \ref{fig::eiger_flux}. The solid circles describe the muon flux detected by the TA station that only traverses the bedrock. The crosses show the muon flux detected by the TB station that traverses the glacier/bedrock. 

We estimated the bulk density of the Eiger bedrock and ice from the measured muon flux shown in Figure \ref{fig::eiger_flux}. We used the TA-measured muon flux (0.846\,cm$^{-2}$sr$^{-1}$day$^{-1}$) crossing 200\,m bedrock at 45\,deg zenith for calibrating our model ($\alpha \sim 3.6$). The calibration is necessary because we ignored the measured open-sky muon flux at the observation place.

Eiger Glacier data was split into four groups: bedrock, bedrock/ice at 60\,deg zenith, bedrock/ice at 70\,deg zenith, and bedrock/ice at 75\,deg zenith. We estimated an average density of about 2.56\,gcm$^{-3}$ for the bedrock data. Then, the estimated density decreases as the observation zenith angle increases, 2.37\,gcm$^{-3}$ for 60\,deg, 1.46\,gcm$^{-3}$ for 70\,deg, and 1.15\,gcm$^{-3}$ for 75\,deg. This tendency indicates that the muon path length is made up of a portion of bedrock and a portion of ice. The authors proposed a model from which we can estimate the bedrock and glacier dimensions along the path length.

\begin{equation}
    <\rho> = x \rho_{rock} + (1-x) \rho_{ice}
\end{equation}
where $x$ is the bedrock portion. Then, $L_{rock} = Lx$ and $L_{ice} = L (1-x)$. Table \ref{tab:table2} summaries the estimated glacier/bedrock densities.

\begin{table}[h!]
\caption{\label{tab:table2}Density estimation}
\begin{ruledtabular}
\begin{tabular}{lccc}
Material & Estimated density [gcm$^{-3}$] & Expected density [gcm$^{-3}$]  & Estimated/Expected ratio  \\ \hline
Khufu pyramid rock    & 2.93 & 2.65  & 1.10  \\ \hline
Canfranc laboratory bedrock & 2.69 & 2.73  & 0.99 \\  \hline
Eiger bedrock      & 2.56 & 2.68 & 0.95 \\ \hline
Eiger bedrock + glacier (60\,deg)    & 2.37 & 2.56\footnotemark[1]  & 0.92  \\ \hline
Eiger bedrock + glacier (70\,deg)    & 1.46 & 2.56  & 0.57  \\ \hline
Eiger bedrock + glacier (75\,deg)    & 1.15 & 2.56  & 0.44  \\ 
\end{tabular}
\end{ruledtabular}
\footnotetext[1]{Assuming as reference the estimated bedrock density.}
\end{table}

\section{Conclusions}
In this work, we presented a methodology for estimating the average density of the bulk material of a target scanned using muography. The method estimated the average density of six different simulated phantoms of toluene, paraffin, water, standard rock, aluminum oxide, and iron. The prediction error was below $10\%$. Furthermore, we performed a validation with real muography data from three independent campaigns, the Khufu pyramid, the Canfranc underground laboratory bedrock, and the Eiger glacier. The model predicted a density of 2.93\,gcm$^{-3}$ for the pyramid rock, 2.69\,gcm$^{-3}$ for the Canfranc laboratory bedrock and 2.56\,gcm$^{-3}$ for the bedrock of the Eiger glacier with a prediction error $<10\%$. This approach also determined a density reduction in the Eiger glacier where a interface ice/bedrock is expected, helping to estimate the dimensions of the glacier as well as the bedrock formation.

% \begin{acknowledgments}

% \end{acknowledgments}

\section*{Data Availability Statement}
The energy loss data used for the phantom simulations are reported by Groom\cite{GROOM2001}. The MUYSC code for simulating the muon flux is available on GitHub\cite{PeaRodrguez2024}.

\section*{References}
\nocite{*}
\bibliography{aipsamp}% Produces the bibliography via BibTeX.

\end{document}